\documentclass[aps,twocolumn,amsmath,amssymb,superscriptaddress,floatfix]{revtex4}
\usepackage{epsf}
\usepackage{psfrag}
\usepackage{graphicx}
\usepackage{color}
\begin{document}

\title{Carbon nanotubes as tunable Luttinger liquids}

\author{Wade DeGottardi}
\affiliation{Department of Physics, University of Illinois at
Urbana-Champaign, 1110 W.\ Green St.\ , Urbana, IL  61801-3080, USA}
\author{Tzu-Chieh Wei}
\affiliation{
    Institute for Quantum Computing and
    Department of Physics and Astronomy,
    University of Waterloo,
    200 University Avenue West,
    Ontario, Canada N2L 3G1}
\author{Smitha Vishveshwara}
\affiliation{Department of Physics, University of Illinois at
Urbana-Champaign, 1110 W.\ Green St.\ , Urbana, IL  61801-3080, USA}

\date{\today}

\title{Transverse field-induced effects in carbon nanotubes}

\begin{abstract}
We investigate the properties of conduction electrons in single-walled armchair
carbon nanotubes (SWNT) in the presence of both transverse electric and
magnetic fields. We find that these fields provide a controlled means of tuning
low-energy band structure properties such as inducing gaps in the spectrum,
breaking various symmetries and altering the Fermi velocities. We show that the
fields can strongly affect electron-electron interaction, yielding tunable
Luttinger liquid physics, the possibility of spin-charge-band separation, and a
competition between spin-density-wave and charge-density-wave order. For short
tubes, the fields can alter boundary conditions and associated single-particle
level spacings as well as quantum dot behavior.
\end{abstract}

\maketitle

\section{Introduction}
The astounding range of experimental and theoretical studies performed on
carbon nanotubes~\cite{Iijima91} has revealed a spectrum of physics
characteristic of strongly correlated low-dimensional electronic
systems~\cite{giamarchi}. The underlying graphene lattice structure of these
tubes uniquely affects band structure, effective dimensionality, Coulomb
interaction effects, and the quantum dot behavior exhibited by short nanotube
segments. The band structure shows differing behavior depending on various
factors such as chirality, applied gate potentials, boundary conditions at the
tube ends and mechanical stress~\cite{ando,saito}. In single-walled armchair
nanotubes (SWNT)~\cite{IijimaIchihashi93}, which are the entities of interest
here, gapless linearly dispersing modes endow the nanotube with its peculiar
quantum wire properties. As described theoretically and ascertained
experimentally, interactions within the modes of this effectively
one-dimensional system cause it to behave as a Luttinger liquid characterized
by non-Ohmic conductances~\cite{kbf,egger,mceuen,dekker,giamarchi}. Tubes
placed between tunnel barriers act as quantum dots~\cite{bockrath,coskun},
which, while displaying zero dimensional physics such as Coulomb blockade
behavior, retain some higher dimensional traits such as hosting plasmons
typical of one dimension and band degrees of freedom attributed to the
underlying graphene lattice. Potentially invaluable to applications, these
nanotube quantum dots have been proposed as elements of quantum devices and
the quantum states of blockaded electrons have been regarded as candidates for
units of quantum information~\cite{qdcnt}. In each of these aspects, the
presence of applied fields can dramatically alter the nanotube's behavior;
here we present an extensive study of the effects of electric and magnetic
fields applied transversally to the axis of the nanotube.

At the level of the band structure, it has been shown that a
parallel magnetic field can have the
 striking effect of converting a metallic tube to a semiconducting one by way of
 inducing a gap~\cite{ando2}, and vice-versa, an effect discernible in conductance, Coulomb
 blockade and scanning tunneling microscope (STM) measurements. Here, instead of a parallel field,
  we discuss
 transverse field configurations (both electric and magnetic) and the conditions under which
a band gap opens up or the spectrum remains gapless in armchair SWNTs. In the
latter case, we demonstrate, via band-structure calculations, simultaneous
breaking of the valley degeneracy (of the two distinct Dirac points), the
left-right-mover degeneracy, and the particle-hole symmetry. Moreover, the
fields yield a non-negligible reduction in the Fermi velocity of conduction
electrons traveling along the tube. We show that for certain configurations of
fields, the ground state of the tube can even be made to carry finite current.

Transverse fields provide an excellent means of altering the ratio of interaction
strength to the Fermi energy in SWNTs. This makes nanotubes potentially the only systems
to date in which the associated Luttinger liquid physics can be tuned in a controlled fashion. As
  described in previous work, either an electric field~\cite{novikov}
  or a magnetic field~\cite{lee,bellucci1} alone suffices to change the value of the Luttinger liquid parameter
from that measured in field-free environments. The  magnitude of the electric
  fields required to bring about a significant change are well within current
  experimental reach \cite{mason}. Here, we find that our approach
  reproduces these results. We show that such a tuning of Luttinger parameters can mediate a transition
  from the system showing tendencies towards spin-density-wave (SDW)
  ordering to that of charge-density-wave (CDW) ordering. Furthermore, in addition to
  the tuning of the Luttinger parameter presented in previous works for the
  net charge density~\cite{novikov,lee,bellucci1}, we find that Luttinger-type interactions become
  manifest in modes associated with the density differences between nanotube
  bands as well. Thus we predict that akin to spin-charge separation, transverse
  fields can induce a spin-charge-band separation wherein the three degrees of
  freedom move at different velocities.

The above results are discussed in the case of an infinite system.
For short tubes or finite length segments formed by tunnel
barriers, boundary effects need to be taken into account.
We find that applied fields influence multiple
aspects of short nanotubes. First, fields can alter the single-particle energy level spacing of the
tube. Here, we carefully account for the effect of the tube ends in the case that the left and right movers travel at different speeds. Second, the charging energies become field dependent, and
third, the plasmon spectrum varies in accordance with the first two effects.
We have conducted a comprehensive analysis of the short nanotube as a
finite-sized Luttinger liquid and show how all three effects can be captured.
Within this description, we discuss the structure of field-dependent Coulomb
blockade peaks and how the presence of both electric and magnetic fields acts
as a means of manipulating quantum states of the dot (the effect of a magnetic field alone
has recently been discussed by Bellucci and Onorato~\cite{bellucci2}).

\begin{figure}
     \includegraphics[width=8.0 cm]{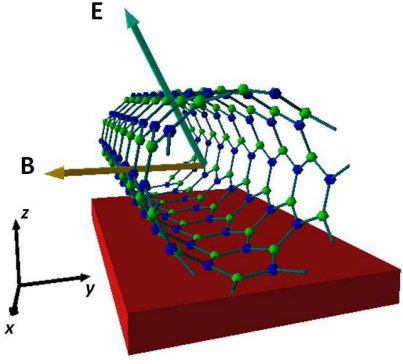}
     \caption{A (5,5) carbon nanotube in the presence of transverse magnetic (pointing in the $-\hat{y}$) and electric fields. The carbon atoms belonging to the A
     and B sublattices are indicated by dark (blue) and light (green) shading, respectively.}
     \label{fig:setup}
\end{figure}

The outline of this paper is as follows. In
section II we present the formulation and results of our band structure
calculation. In section III we formulate an effective one-dimensional
Hamiltonian which takes into account field effects. We bosonize this
Hamiltonian, describing interaction effects in terms of Luttinger liquid
physics. In section IV we investigate the various Luttinger liquid phases and
the feasibility of using fields to access such phases. In section V we discuss
field-tuned quantum dot physics. Finally, in section VI we present the
highlights of our results and discuss their relevance to experiments.

\section{Band structure in transverse fields}

We briefly recapitulate the band structure of an infinitely long armchair tube
in the absence of any fields~\cite{saito}. The electronic properties of
graphene are well described by a tight-binding model in which electrons hop
between nearest neighbors of the underlying hexagonal bipartite lattice
(sublattices here labeled $A$ and $B$) with an associated energy (the hopping
integral) of $t\approx 3$ eV. An armchair carbon nanotube can be regarded as a
sheet of graphene rolled along the $(n,n)$-direction (for notation of the
chirality, see e.g. Ref.~\cite{saito}), denoted by $\hat{s}$. This gives rise
to states of quantized momentum $k_s = (0, 2 \pi/L, ... , 2 \pi(2n-1)/L)$
where $L = \sqrt{3} n a$ is the circumference of the tube, and $a = \sqrt{3}
a_{c}$ where $a_{c} \approx 0.15$ nm is the nearest carbon-carbon distance.
The resulting series of one dimensional bands can be described by the
wavevectors $\vec{k} = (k_x, \frac{2 \pi \ell}{L})$ where $k_x$ is the
quasimomentum parallel to the tube's axis and $\hbar \ell$ is the state's
angular momentum about the tube's circumference. A convenient set of basis
states is given by the following linear combination of atomic orbitals
\begin{equation}
| \Phi_{A/B}^\ell \rangle = \frac{1}{\sqrt{2n}} \sum_{\vec{R} \in A/B} e^{i \vec{k} \cdot \vec{R} } | \vec{R} \rangle,
\end{equation}
where $| \vec{R} \rangle$ is the $\pi$-electronic state of the atom located at
$\vec{R}$. The sum runs over the $n$ atoms in the unit cell that belong to
either the A or B sublattice. At half filling, the associated dispersion has
low energy excitations near the so-called Dirac points of the form $\epsilon=
\pm \hbar v_F|k - \alpha k_F |$ where $v_F= \frac{\sqrt{3} t a}{2 \hbar}
\approx 8 \times 10^5$ m/s, $k_F = 4 \pi/3 a$ and $\alpha = \pm$. Thus,
$k=\alpha k_F$ label the two inequivalent Fermi points.

The setup of interest is shown in  Fig.~\ref{fig:setup}. An external magnetic
field is applied in the negative $y$-direction; an applied (transverse)
electric field makes an angle $\chi$ with the magnetic field. These fields
give rise to scalar and vector potentials
\begin{eqnarray}
U(s) &=& |e| R \cos \left( \frac{s}{R} - \chi \right) \\
\vec{A} &=& - B z \hat{x} \label{eq:mag},
\end{eqnarray}
respectively ($R = L/2\pi$). The $x$-axis runs parallel to the tube's axis and
the additional coordinate $s$ measures the circumferential distance starting
from the negative $y$-axis (a positive value of $s$ corresponds to a
counterclockwise rotation as one looks along the $x$-axis in the positive
direction).

These external potentials are easily accommodated within the tight binding
approach. In Eq.~(\ref{eq:mag}) we have selected a gauge that is independent
of $x$ and thus $k_x$ remains a good quantum number. The hopping matrix
elements in the presence of the fields are given by
\begin{equation}
\label{eq:magmatrix}
\langle \Phi^{\ell'}_A \left| H \right| \Phi^{\ell}_B \rangle = -\frac{t}{2n} \sum_{\vec{R} \in B, \vec{R}' \in A} e^{i(k \cdot R - k' \cdot R') + \frac{i e}{\hbar}(G_R-G_R')}
\end{equation}
where the sum runs over nearest neighbors $\vec{R}$, $\vec{R}'$ and
\begin{equation}
G_j - G_i \approx \int_{0}^{1} d\lambda \ (\vec{r}_i - \vec{r}_j) \cdot \vec{A}\left(\vec{r} + \lambda(\vec{r}_i - \vec{r}_j)\right)
\end{equation}
is the Aharonov-Bohm phase associated with the magnetic field \cite{saito}. The
dimensionless parameter $b$ is given by $b = B \frac{ \sqrt{3} |e| L^2}{4 \pi^2
\hbar}$. Numerically, for an $(n,n)$ nanotube, the magnetic field in Teslas is related to the dimensionless parameter $b$ via $B \approx 8.1 \times 10^4 \times b /n^2$ T. The scalar potential gives rise to an on-site potential described by
the matrix element
\begin{eqnarray}
\label{eq:elmatrix}
\langle \Phi^{\ell'}_A \left| H \right| \Phi^{\ell}_A \rangle = \langle \Phi^{\ell'}_B \left| H \right| \Phi^{\ell}_B \rangle = \frac{U}{2} e^{\pm i \chi}
\end{eqnarray}
for $\ell' = \ell \mp 1 \mod n$ where $U = |e| E R / t$. The electric field
strength in V/nm is related to $U$ by $E \approx 42 \ U/n$ V/nm for a tube with
chiral vector $(n,n)$. That these matrix elements mix states of different
angular momentum has a straightforward classical analog: a charged particle on
the surface of a cylinder will circulate around its circumference as a result
of the applied fields

\begin{figure}
    \includegraphics[width=8.0 cm]{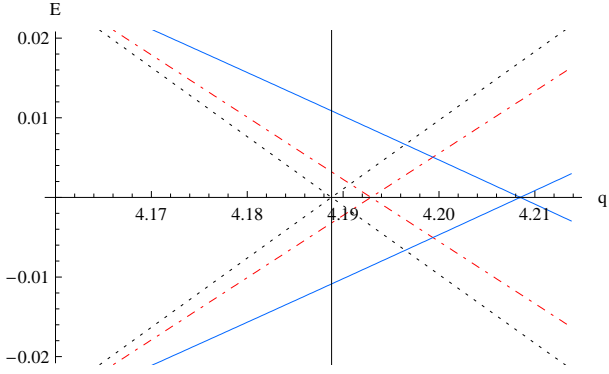}
    \caption{ Spectrum of a $(5,5)$ carbon nanotube near the
    $\alpha=+$ Dirac Fermi point (field-free value $k_F = \alpha 4\pi/3a$ indicated by the vertical line) in the presence of an external perpendicular field with $U_0/t = 0$ (black dotted), $0.2$ (red dot-dashed), $0.4$ (blue solid)
    as the crossing moves to the right. The horizontal axis indicates the value of $q$, where $q = ka$; the vertical axis is given in units of $t$, the hopping integral ($t \approx 3 $ eV). \label{fig:spectrumE}}
\end{figure}

\begin{figure}
     \includegraphics[width=8.0 cm]{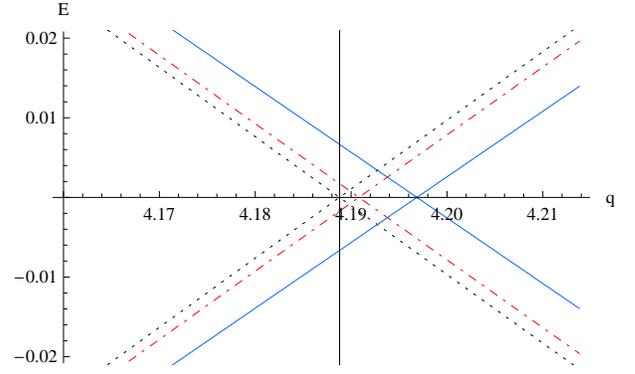}
     \caption{Spectrum of a $(5,5)$ carbon nanotube near the
    $\alpha=+$ Dirac Fermi point (field-free value $k_F = \alpha 4\pi/3a$ indicated by the vertical line) in the presence of a magnetic field $b = 0$ (black dotted), $0.2$ (red dot-dashed), $0.4$ (blue solid)
     (as the crossings move to the right). The horizontal axis indicates the value of $q$, where $q = ka$; the vertical axis is given in units of $t$, the hopping integral ($t \approx 3 $ eV). \label{fig:spectrumB}}
\end{figure}

\begin{figure}
    \includegraphics[width=8.0 cm]{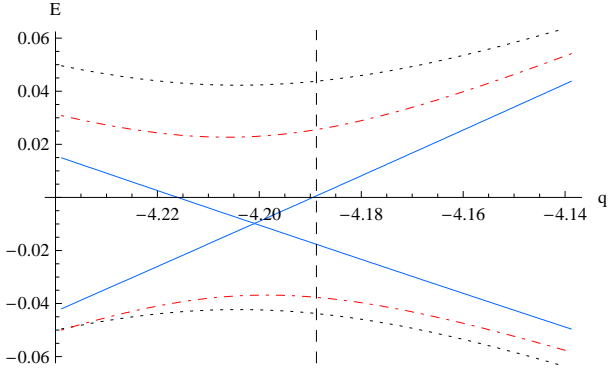}
    \caption{Spectrum of a $(5,5)$ carbon nanotube near $\alpha=-$ Dirac Fermi point (field-free value $\alpha k_F = -4\pi/3a$ indicated by the vertical dashed line)
    in the presence of transverse electric and magnetic field ($U/t = 0.2$ and $b = 0.4$). The angle between $\vec{E}$ and $\vec{B}$ being $0$ (black dotted), $\pi/4$ (red dot-dashed), $\pi/2$ (blue solid) (from outer to inner). The horizontal axis indicates the value of $q$, where $q = ka$; the vertical axis is given in units of $t$, the hopping integral ($t \approx 3 $ eV).
    \label{fig:spectrumVBN}}
\end{figure}

\begin{figure}
     \includegraphics[width=8.0 cm]{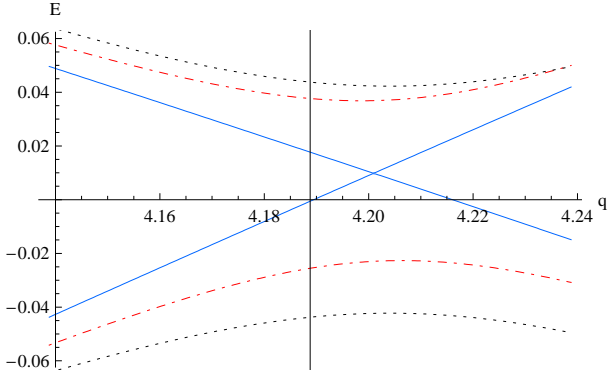}
     \caption{Spectrum of a $(5,5)$ carbon nanotube near $\alpha=+$ Dirac Fermi point (field-free value $\alpha k_F = 4\pi/3a$ by the vertical  line) in the presence of transverse electric and magnetic field ($U/t = 0.2$ and $b = 0.4$). The angle between $\vec{E}$ and $\vec{B}$ being $0$ (black dotted), $\pi/4$ (red dot-dashed), $\pi/2$ (blue solid) (from outer to inner). The horizontal axis indicates the value of $q$, where $q = ka$; the vertical axis is given in units of $t$, the hopping integral ($t \approx 3 $ eV).
    \label{fig:spectrumVB}}
\end{figure}

We have studied the effects of the fields perturbatively in $b$ and $U$. In
the vicinity of the Fermi points, the left and right moving bands are nearly
degenerate, so care must be taken in applying perturbation theory. The details
of this calculation are presented in Appendix~\ref{app:band} where we carry
out perturbation theory to second order. There are three cases of interest
that we summarize below. An illustration of these three cases is shown in
Figs.~\ref{fig:spectrumE},~\ref{fig:spectrumB},~\ref{fig:spectrumVBN},
and~\ref{fig:spectrumVB}. While some of the effects discussed are rather small
for standard SWNTs, we note that our band structure analysis can be applied to
multi-walled nanotubes as well in which case a larger radius yields more
pronounced effects.

\subsubsection{Case of $E = 0$ or $B = 0$}

For a single field, the most salient features of our band structure calculation
are the reduction in Fermi velocity and the shift in Fermi momentum.
Semiclassically, the reduction in Fermi velocity can be ascribed to the
deflection of the electrons by the fields leading to a reduction in the
velocity component along the tube. Furthermore the bands remain gapless.

For a magnetic field, we find a reduced Fermi velocity given by
\begin{equation}
\tilde{v}_F = v_F \left(1 - \Delta v_1 b^2 \right).
\end{equation}
The first order correction in $b$ vanishes because $\tilde{v}_F$ must be an
even function of $b$ by symmetry. The term $\Delta v_1$ is a function of $n$
and is given by Eq.~(\ref{eq:v1}). This term depends on the geometric details
of an armchair tube; for large tubes $\Delta v_1 \approx 1/3$. An experimental
observation of such a reduction will require strong fields. For example, a
(20,20) armchair tube with a 20 T field gives a $0.4\%$ reduction in the Fermi
velocity.

For an electric field we have
\begin{equation}
\tilde{v}_F = v_F \left(1 - \Delta v_2  U^2 \right).
\end{equation}
A field of strength $0.1$ V/nm corresponds to a reduction in the Fermi velocity
of a (10,10) tube of roughly $10\%$. The term $\Delta v_2 \approx n^2/\pi^2$
for large tubes; its exact form is given by Eq.~(\ref{eq:v2}).

As mentioned in the introduction, the reduction of the Fermi velocity has been
noted by several authors~\cite{novikov,lee,bellucci1}. In these papers the
carbon nanotube was modeled as a smooth cylinder and the low-energy electronic
behavior was put in by hand. Our results are in agreement with these results in
the limit of small fields and large $n$ (in the regime that perturbation is
valid). Additionally, by taking into account the geometry of the armchair
nanotube we find that either a magnetic or electric field alone will shift the
Fermi points. That is, the nanotube still has the same low energy spectrum with
renormalized values of $k_F$ (whose precise form is given by
Eq.~(\ref{eq:kf})). The band structure of a $(5,5)$ nanotube in the presence of electric and magnetic fields of various strength is shown in Figs.~\ref{fig:spectrumE} and~\ref{fig:spectrumB}, respectively.

\subsubsection{Case of $\vec{E} \bot \vec{B}$ ($\chi=\pi/2$)}

Mutually perpendicular fields break both the time-reversal and particle-hole
symmetry of the band structure. The left and right movers now move with
different speeds. For a magnetic field in the negative $y$-direction and an electric
field in the positive $z$-direction we have
\begin{equation}
\tilde{v}_{r} = v_F \left(1 - \Delta v_1 b^2 - \Delta v_2 U_z^2 \pm \Delta v_3 b U_z \right),
\end{equation}
where $r$ is $+$ for right movers and $-$ for left movers. The expressions for
$\Delta v_3$ is given by Eq.~(\ref{eq:v3}); for large tubes $\Delta v_3 \approx
n /\pi $. For the fields we consider, $E/B$ is roughly the same order of
magnitude as $v_F$ and thus it is natural to expect the band structure will
mimic the behavior of a classical velocity selector. Indeed the asymmetry of
the velocity in the right- and left-moving branches is expected from a
elementary consideration. For a charge particle moving on a smooth cylinder in
the presence of mutually perpendicular transverse magnetic and electric fields,
the direction of the force caused by the magnetic field depends on whether the
particle is moving along the tube in one direction or another, whereas the
electric field remains the same. Hence, for fixed kinetic energy, the force in
the transverse direction causing the particle to spiral is different for
different directions of motion, and giving rise to different velocities along
the axial direction.

Another prominent feature of the band structure in this case is a relative
energy shift with respect to the two Fermi points. Near the two Dirac points
$k_F \approx \pm \frac{4 \pi}{3 a}$ we have
\begin{equation}
\label{eq:mdiv} \epsilon_{r \alpha} (k) =  \hbar r \tilde{v}_{r} \left( k -
\alpha \tilde{k}_F \right) + \alpha t \Delta s + \mathcal{O}(\Delta k^2),
\end{equation}
where $\alpha$ ($\alpha = \pm$ for $k_F \approx \pm \frac{4\pi}{3a}$). By definition, $\tilde{k}_F$ is the momentum for which the left and right moving bands for a given Fermi point are degenerate and again is generally different from $4 \pi/3a$ in the presence of fields. The precise forms of $k_F$ and $\Delta s$ are given by Eqs.~(\ref{eq:kf}) and (\ref{eq:ds}), respectively. The solid blue lines in Figs.~\ref{fig:spectrumVBN} (for $\alpha = +$) and~\ref{fig:spectrumVB} ($\alpha = -$) indicate low-energy dispersion of a $(5,5)$ nanotube in the presence of crossed electric ($v = 0.2$) and magnetic ($b = 0.4$) fields.

\subsubsection{Case of $\chi \neq \frac{\pi}{2}$}
The presence of both electric and magnetic fields will generically open gaps at
the Fermi points. The size of this gap is given by
\begin{equation}
\epsilon_{gap} \approx t b U \frac{2 \sqrt{3}\cos \frac{\pi}{3n}}{1 + 2 \cos{\frac{\pi}{3n}}} \left| \cos \chi \right|. \label{eq:gap}
\end{equation}
For example, a (15,15) tube parallel electric and magnetic fields 1 V/nm and
10 T respectively gives $\epsilon_{gap} / k_B \approx  4$ meV. For electric
and magnetic fields which are not parallel, this gap is weakly indirect.

The degeneracy associated with the graphene Fermi points stems from the
equivalence of the two sublattices. For example, a two-dimensional honeycomb
lattice with A and B sublattices composed of different types of atoms would
generically have a gap \cite{saito}. Now, in a nanotube, gaps can arise for
different reasons. For example, the gap associated with a semiconducting tube
occurs because the quantized bands miss the Dirac point of the underlying
graphene lattice. Such a gap can be closed by applying a magnetic field (of a
specific strength) parallel to the tube's axis. In the present case however,
the gap arises from an energy difference associated with the two sublattices
and therefore can not be so closed. Not surprisingly, the gap described here
vanishes precisely when the electric and magnetic forces on a classical charged
particle traveling along the tube are either parallel or antiparallel.

From Figs.~\ref{fig:spectrumVBN} and~\ref{fig:spectrumVB} we clearly see
that as the angle between the fields varies, the particle-hole symmetry and
the valley degeneracy, as well as the symmetry between left- and right-moving
spectra, are broken. The degree of breaking of these symmetries is greatest
in the case that the fields are perpendicular. On the other hand, these symmetries are
preserved when the fields are parallel or anti-parallel and the gap is at its
maximum. This angle-dependent gap can be manifest in the transport
measurement, e.g., the conductance can be $4e^2/h$, or $2e^2/h$ or zero,
depending on whether there are $4$, $2$ or zero conducting channels (including
spin).

\subsubsection{Low-energy modes}
In order to study the low-energy physics of electronic excitations about the
Fermi energy for all the cases described above, we can identify the fermionic
operator on the cylindrical surface of the tube as approximately
\begin{equation}
\Psi_{\sigma}(x,s) = \sum_{p \alpha} \varphi_{p \alpha}(x,s) \psi_{p \alpha
\sigma}(x), \label{eq:wvfn}
\end{equation}
where $\psi_{p \alpha \sigma}(x)$ is the one-dimensional field operator at the
point $x$ along the tube axis associated with the A and B sublattices ($p =
A(+)/B(-)$), Fermi points $k_F = \alpha 4\pi/3a$ ($\alpha = \pm$), and spin
$\sigma = \uparrow(+)/\downarrow(-)$. The Bloch wavefunction $\varphi_{p
\alpha}(x,s)$ retains detailed information about the response of the
electronic wave functions to the applied fields \cite{egger,kbf}. While the
sublattice basis (indexed by $p$) is convenient for discussing these
wavefunctions, it does not diagonalize the hopping interaction. We therefore
transform to the right and left moving basis ($r = R(+)/L(-)$) via the
transformation $\psi_{p \alpha \sigma} = \sum_r U_{p r} \psi_{r \alpha
\sigma}$ where $U^\dagger \sigma_y U = \sigma_z$ \cite{egger}. This gives a
kinetic energy term
\begin{equation}
H_0=- i \hbar \sum_{r \alpha} \int dx \ r v_r \psi_{r \alpha}^\dagger
\partial_x \psi_{r \alpha}, \label{eq:hkinetic}
\end{equation}
where, as shown above a generic field configuration can give rise to the
possibility that $v_R \neq v_L$.

The gap which arises as a result of the presence of both electric and magnetic
fields [Eq.~(\ref{eq:gap})] can be incorporated into our Hamiltonian by a mass
term
\begin{equation}
H_{\rm gap1}=\sum_{r \alpha \sigma} \int dx \ \frac{\epsilon_{gap}}{2}
\psi^\dagger_{r \alpha \sigma} \psi_{-r \alpha \sigma}. \label{eq:gap1}
\end{equation}
Note that this mass term is of a different form than that considered in the
exhaustive zero-field study by Egger et al.\cite{egger} (which had the form $i
\frac{\epsilon_{gap}}{2} r \psi^\dagger_{r \alpha \sigma} \psi_{-r \alpha
\sigma}$). Similarly, the effect of the shift associated with the Fermi points
can be described by the term
\begin{equation}
H_{\rm gap2}= \sum_{r \alpha \sigma} \int dx \ \alpha\frac{t\Delta s}{2}
\psi^\dagger_{r \alpha \sigma} \psi_{r \alpha \sigma}.\label{eq:gap2}
\end{equation}

\section{Luttinger liquid formulation}

In this section we specialize to the case of gapless modes for which
interactions can be easily incorporated. As the simplest case, when only a
magnetic or electric field is present, the kinetic piece of the Hamiltonian
given in Eq.~(\ref{eq:hkinetic}) now has $\tilde{v}_{+} =
\tilde{v}_{-}\equiv\tilde{v}_F$ and $\epsilon_{gap} = 0$, i.e.,
\begin{equation}
H_0 = - i \hbar \tilde{v}_F \sum_{r \alpha \sigma} \int dx \ \psi^\dagger_{r
\alpha \sigma} \partial_x \psi_{r \alpha \sigma}. \label{eq:1dKE2}
\end{equation}
We use standard approaches such as bosonization to study the effect of fields.
Where appropriate, we include discussions for the asymmetric case of
$\tilde{v}_{+}\neq \tilde{v}_{-}$. We closely follow the approach of
Ref.~\cite{egger} whose lucid pedagogical exposition we do not repeat but
instead confine our discussion to the new field-dependant features.

The presence of fields does not alter the fact that the electrons on a tube are
locked into their lowest energy radial mode. Hence, it is possible to study the
low energy excitations in the presence of interactions using an effective
bosonized 1D Hamiltonian. Bosonization offers a great simplification since many
of the quartic interaction terms in the fermionic language become quadratic
once they are bosonized.

The interaction term takes the general form
\begin{equation}
H_{int} = \frac{1}{2} \int d \mathbf{r} \int d \mathbf{r}' \
\Psi_\sigma^\dagger(\mathbf{r}) \Psi_{\sigma'}^\dagger(\mathbf{r}')
U(\mathbf{r}-\mathbf{r}') \Psi_{\sigma'}(\mathbf{r}') \Psi_\sigma(\mathbf{r}),
\label{eq:Hint1}
\end{equation}
where $\Psi_\sigma(\mathbf{r})$ is the field in Eq.~(\ref{eq:wvfn}) describing
low-energy electrons.

Following \cite{egger}, we employ the form of the Coulomb interaction on the
surface of a cylinder given by
\begin{equation}
U(x - x', s - s') = \frac{e^2/\kappa}{\sqrt{(x-x')^2 + 4 R^2 \sin^2
\left(\frac{s - s'}{2R}\right) + a_z^2}}, \label{eq:coulomb}
\end{equation}
where $R$ is the radius of the tube and $x$ and $s$ the coordinates defined in the previous section and $a_z \approx a$ is roughly the thickness of the graphene sheet ~\cite{egger}. The form of the interaction in Eq.~(\ref{eq:Hint1}) is explicitly given in terms of two-dimensional integrals. An
effective interaction term involving purely linear integrals along the tube's
axis can be obtained by expressing $\Psi_\sigma(\mathbf{r})$ in terms of the
linear and circumferential fields as in Eq.~(\ref{eq:wvfn}) and integrating out
the circumferential degrees of freedom from Eq.~(\ref{eq:Hint1}). Here we
remark that integration over the circumferential degrees of freedom generates
field-induced terms that give rise to novel physics and can be traced to the
dependence of the ground state wave function on non-zero angular momentum states
in the circumferential direction.

The resulting effective interaction involves two-particle scattering processes
between fermions moving along the tube axis denoted by the fields $\psi_{r
\alpha \sigma}$. The associated scattering processes can be classified
according to whether the incoming particles preserve their Fermi point quantum
number $\alpha$ when scattered, - forward scattering ($\alpha FS$)-, or
scatter across the Fermi surface - backscattering ($\alpha BS$). In Ref.
\cite{egger}, a further distinction is made in the forward scattering events
depending on whether the interaction potential is homogeneous over the
circumference of the tube ($\alpha FS0$) or is able to distinguish the
microscopic differences which arise between sublattices ($\alpha FS1$). At
this point, in order to obtain an effective low-energy description of the
interacting electrons, the one-dimensional fermionic operators can be
bosonized as in Ref.~\cite{egger}:
\begin{equation}
\psi_{r \alpha \sigma} = \frac{\eta_{r \alpha \sigma}}{\sqrt{2 \pi a_{c}}} \exp \left[ i \alpha k_F x + i \varphi_{r \alpha \sigma} \right].
\end{equation}
where
\begin{eqnarray}
\varphi_{r \alpha \sigma} &=& \frac{\sqrt{\pi}}{2} ( \phi_{c+} + r \theta_{c+}
+ \alpha \phi_{c-} + r \alpha \theta_{c-} \\ \nonumber
 &+& \sigma \phi_{s+} +
r \sigma \theta_{s+} + \alpha \sigma \phi_{c-} + r \alpha \sigma \theta_{s-}).
\label{eq:basisdef}
\end{eqnarray}
The bosonic fields $\varphi$'s satisfy the commutation relations
\begin{equation}
\label{eq:comm}
[\varphi_{r\alpha\sigma}(x),\varphi_{r'\alpha'\sigma'}(x')]=-i\pi
r\delta_{rr'}\delta_{\sigma\sigma'}{\rm sgn}(x-x'),
\end{equation}
where $r=\pm$ denotes the left- and right-movers, $\alpha=\pm$ indicates the
Fermi points, and $\sigma=\pm$ represents the spin direction
($\uparrow/\downarrow$). The  $\eta_{r \alpha \sigma}$'s are the so-called
Klein factors; they enforce the anticommutation relations between different
channels,
\begin{equation}
\{\eta_{r\alpha\sigma},\eta_{r'\alpha'\sigma'}\}=2\delta_{rr'}\delta_{\alpha\alpha'}\delta_{\sigma\sigma'}.
\end{equation}
Moreover, the fields $\theta_{j\delta}(x)$'s (with $j=c/s$ and $\delta=\pm$)
and their dual fields $\phi_{j\delta}(x)$ [both are linear combinations of
$\varphi_{r\alpha\sigma}(x)$] in turn satisfy
\begin{equation}
[\phi_{j\delta}(x),\theta_{j'\delta'}(x')]=-\frac{i}{2}
\delta_{jj'}\delta_{\delta\delta'}{\rm sgn}(x-x')
\end{equation}
 The effective density in a given channel takes the form
\begin{equation}
\widetilde{\rho}_{r \alpha \sigma}(x) = \frac{r}{2 \pi} \partial_x \varphi_{r \alpha \sigma}(x).
\label{eq:bosondensity}
\end{equation}
The kinetic energy associated with the linearly dispersing fermionic modes is
quadratic in the bosonized fields. As for the interactions, the dominant
contributions also come from quadratic terms reflecting net density-density
type interactions. As in the field-free case, the $\alpha FS0$ process has one
such contribution from the usual Coulomb form involving the net charge
density, which in the bosonized representation is given by
\begin{equation}
H_{\alpha FS,0} = \frac{2}{\pi} \int dx \ \widetilde{V}(k \approx 0) \left(
\partial_x \theta_{c+} \right)^2,
\label{eq:FS0}
\end{equation}
where
\begin{equation}
\widetilde{V}(k) \approx \frac{2 e^2}{\kappa} \left( |\ln{k R}| + c_0 \right)
\label{eq:one}
\end{equation}
is the Fourier transform of $V(x)$ and $c_0$ is a function of $n$ (see Eq.
(\ref{eq:c})).

The presence of either an electric or magnetic field gives rise to additional
quadratic terms in the $\alpha FS0$ process . These terms have their origin in
the non-zero angular momentum components of the circumferential wavefunction
$\varphi_{r \alpha}(x,s)$ in Eq.~(\ref{eq:wvfn}). A detailed accounting of the
radial wave functions (see appendix B) shows that an electric field
contributes a term
\begin{equation}
H_{\alpha FS,E} = \int dx \left(\frac{2 e^2}{\kappa}\right) \left(\frac{2}{\pi}
\right) h_1 U^2 \left(
\partial_x \phi_{c+} \right)^2
\label{eq:FSE}
\end{equation}
whereas a magnetic field provides a contribution
\begin{equation}
H_{\alpha FS,B} = \int dx \left(\frac{2 e^2}{\kappa}\right) \left(\frac{2}{\pi}
\right) h_2 b^2 \left(\partial_x \theta_{c-} \right)^2. \label{eq:FSB}
\end{equation}
In the presence of mutually perpendicular transverse electric and magnetic
fields, there is an additional contribution to the interaction
\begin{equation}
H_{\alpha FS,BE} = \int dx \left(\frac{2 e^2}{\kappa}\right)
\left(\frac{2}{\pi} \right) h_3 U b (\partial_x \theta_{c-})
(\partial_x\phi_{c+}). \label{eq:FSBE}
\end{equation}
 The values of
$h_1$, $h_2$, and $h_3$ are given in appendix B. The Luttinger liquid
Hamiltonian takes into account all these quadratic terms; other terms emerging
from the interaction potential are sub-dominant and can be considered
perturbatively.

An intuitive physical picture for the origin of the field-dependant
interaction terms above can be obtained by noting that in Eq.~(\ref{eq:rhora})
the charge density for $\rho_{r\alpha}$ is of the form $1+ t_1 r u +t_2 \alpha
b$. This is broadly consistent with the magnetic field coupling to momentum
via $(p-eA/c)^2$ and hence to $\alpha k_F$, resulting in a term $\sim \alpha
b$. Moreover, the electric potential differs slightly between adjacent $A$ and $B$ sites,
and in turn couples to left- and right-movers differently, resulting in a
term $\sim r U$. Equation~(\ref{eq:Hint1}) thus gives rise to interactions among the charge densities
$\partial_x\theta_{c\pm}$ and current $\partial\phi_{c+}$. We therefore obtain the following
nonvanishing terms: (1) $\partial_x\theta_{c+}$, (2)
$\partial_x\theta_{c-}\sim b$, and (3) $\partial_x\phi_{c+}\sim u$. Hence,
one expects the resulting interaction terms given above.

The sum of the kinetic energy and interactions described by
Eqs.~(\ref{eq:1dKE2}),~(\ref{eq:FS0}),~(\ref{eq:FSE}), and~(\ref{eq:FSB}) gives
the total Hamiltonian that we focus on in this paper
\begin{equation}
H_{tot} = H_{0} + H_{\alpha FS,0} + H_{\alpha FS,E} + H_{\alpha FS,B}.
\label{eq:H_tot}
\end{equation}
For the case of a single field (either magnetic or electric) the dispersion
remains symmetric and therefore Eq.~(\ref{eq:H_tot}) can be written in the
form
\begin{equation}
H_{tot} = \sum_{a = \pm c/s} \frac{v_a}{2} \int dx \left[ \frac{1}{K_a}
\left(\partial_x \theta_a \right)^2 + K_a \left( \partial_x \phi_a
\right)^2\right]. \label{eq:LL}
\end{equation}
$K_a=1$ reflects no interactions in the '$a$' sector and $K_a<1$ reflects
repulsive interactions. For the asymmetric case, the diagonalization of the
Luttinger liquid Hamiltonian is technically more complicated but conceptually
simple given the quadratic form of the relevant terms~\cite{trushin}. Below we
discuss the form of the various Luttinger parameters and the related physics.

\subsection{Luttinger parameters for $B = 0$, $E \neq 0$}

\begin{figure}
     \includegraphics[width=8.0 cm]{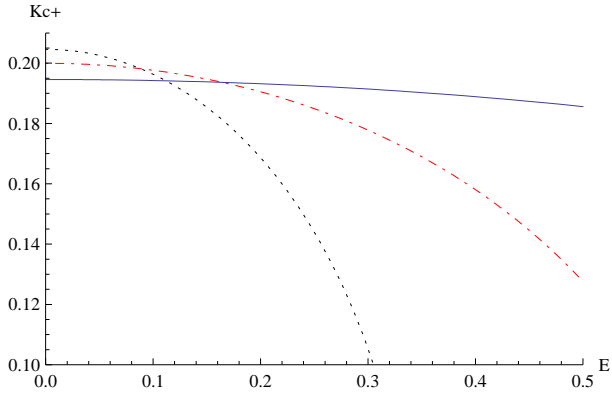}
     \caption{Luttinger parameter $K_{c+}$ of a 500 nm long nanotube as a function of electric field strength $E$ (in V/nm) for different values of $n$: $n = 10$ (solid blue line), $n=15$ (red dot-dashed line), and $n=20$ (black dotted line).
     \label{fig:kcplus}}
\end{figure}

As seen above, for just an electric field present, interactions only affect
the net charge sector $c+$ and renormalize the velocity in this sector
\begin{eqnarray}
v_{c+} K_{c+}&=&\tilde{v}_F+ \frac{8 e^2 h_1 U^2}{\pi \kappa},\\
v_{c+}/K_{c+}&=&\tilde{v}_F+ \frac{8 e^2}{\pi \kappa}( |\ln k_cR | + c_0),
\end{eqnarray}
where $k_c\approx 1/L$ is the lower cutoff provided by the length $L$ of the
tube. This yields a Luttinger parameter value
\begin{equation}
K_{c+} =  \sqrt{\frac{ 1 + \frac{8 e^2 h_1 U^2}{\pi \kappa \tilde{v}_F(U)}}{1
+ \frac{8 e^2}{\pi \kappa \tilde{v}_F(U)} \left( |\ln k_cR | + c_0
\right)\label{eq:KcpE}}}
\end{equation}
in the low-field limit, where $\tilde{v}_F(U)$ is the field reduced Fermi
velocity calculated in the previous section. Figure~\ref{fig:kcplus} shows the
dependence of $K_{c+}$ on $U$. In the other sectors, the Luttinger parameters
retain their non-interacting value $K_{c-} = K_{s\pm}=1$. The Luttinger model
predicts power law behavior for the tunneling density of states~\cite{giamarchi} with exponents $\alpha_{end} = (K_{c+}^{-1}-1)/4$ and
$\alpha_{bulk} = (K_{c+} + K^{-1}_{c+} - 2)/8$ for tunneling into the end or
bulk of a tube respectively.

A smaller value of $K_{c+}$ implies stronger repulsive interactions in the net
charge sector. Here, two tendencies compete; in Eq.~(\ref{eq:KcpE}),
$\tilde{v}_F(U)<{v}_F$ increases the interaction strength relative to the
kinetic energy. On the other hand, the direct effect of the field, as
reflected in the $h_1U^2$ term in Eq.~(\ref{eq:KcpE}) is to decrease the
relative interaction strength. The former effect is dominant of the latter,
hence as shown in Fig.~\ref{fig:kcplus},  $K_{c+}$ is monotonically decreasing
in increasing field strength.

\subsection{Luttinger parameters for $B \neq 0$, $E = 0$}

\begin{figure}
     \includegraphics[width=8.0 cm]{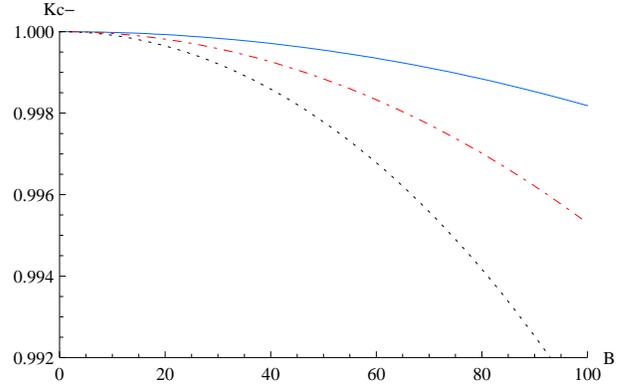}
     \caption{Luttinger parameter $K_{c-}$ of a 500 nm long nanotube as a function of magnetic field strength $B$ (in Teslas) for different values of $n$: $n = 10$ (solid blue line), $n=15$ (red dot-dashed line), and $n=20$ (black dotted line).
     \label{fig:kcminus}}
\end{figure}
In the presence of a magnetic field, the Luttinger parameter deviates from the
non-interacting value of unity not only in the net charge sector,
\begin{equation}
K_{c+}^{-1} = \sqrt{1 + \frac{8 e^2}{\pi \kappa \tilde{v}_F(b)} \left( |\ln
k_cR | + c_0 \right)},
\end{equation}
but also in the relative charge sector,
\begin{equation}
K_{c-}^{-1} = \sqrt{ 1 + \frac{8 e^2 h_2 b^2 }{\pi \kappa \tilde{v}_F(b)}}.
\label{eq:KcmB}
\end{equation}
The spin sectors remain unaffected; $K_{s\pm}=1$.

For both cases $c+$ and $c-$, the reduction in the Fermi velocity
$\tilde{v}_F(b)<v_F$ increases the relative interaction strength, thereby
decreasing the values of $K_{c\pm}$. In addition, as reflected in the $h_2b^2$
coefficient in Eq.~(\ref{eq:KcmB}), the field directly decreases $K_{c-}$ via
the interaction term of Eq.~(\ref{eq:FSB}) discussed above.
Figure~\ref{fig:kcminus} shows the dependence of $K_{c-}$ on the magnetic
field $B$. In this case, the Luttinger model predicts power law behavior for
the tunneling density of states with exponents $\alpha_{end} =
(K^{-1}_{c+}+K^{-1}_{c-}-2)/4$ and $\alpha_{bulk} = (K_{c+} + K^{-1}_{c+} +
K_{c-} + K^{-1}_{c-} - 4)/8$. Unfortunately, this effect is small; even for
fields as large as $100T$, $K_{c-}$ is reduced from unity only by about one
half of a percent for a (20,20) tube (see Fig.~\ref{fig:kcminus}).
Nevertheless, for multi-wall nanotubes, the outermost layer can have $n$ as
large as 100, and hence, the field only needs to be as large as  $4T$ to see a
0.5 \% reduction.

The deviation of $K_{c-}$ from unity leads to the fascinating prospect of
spin-band-charge separation. In one-dimensional systems, the possibility of
spin-charge separation stemming from different interactions within the two
sectors has been extensively discussed and observed in the case of etched
quantum wires~\cite{Yacoby05}. Here we predict that nanotubes in transverse
magnetic fields can undergo yet another separation due to the interactions in
the $c-$ sector. Thus we propose that in this case, the four modes travel at
three different velocities, $v_{c+}=\tilde{v}_{F}(b)/K_{c+}$,
$v_{c-}=\tilde{v}_F(b)/K_{c-}$ and $v_{s\pm}=\tilde{v}_F(b)$.

\subsection{Luttinger parameters for $\vec{E} \bot \vec{B}$}

\begin{figure}
     \includegraphics[width=8.0 cm]{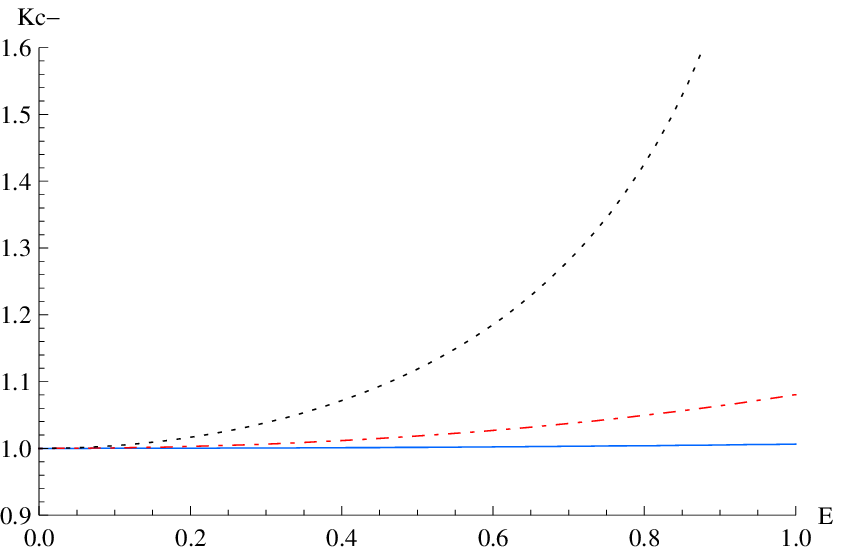}
       \caption{Luttinger parameter $K_{c+}$ of an $n = 10$ (solid blue line), $n=15$ (red dot-dashed line), and $n=20$ (black dotted line) carbon nanotube as a function of electric field strength E (in V/nm) in the presence of a 5T magnetic field (tube has $K^0_{c+} = 0.2$). See Eq.~(\ref{eq:vBE})}
     \label{fig:kcminus2}
\end{figure}

\begin{figure}
     \includegraphics[width=8.0 cm]{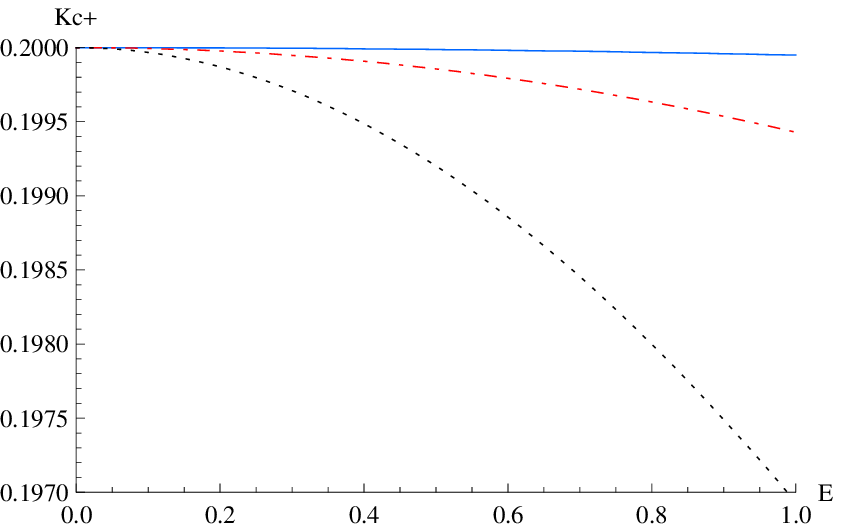}
     \caption{Luttinger parameter $K_{c+}$ of an $n = 10$ (solid blue line), $n=15$ (red dot-dashed line), and $n=20$ (black dotted line) carbon nanotube as a function of electric field strength E (in V/nm) in the presence of a 5T magnetic field (tube has $K^0_{c+} = 0.2$). See Eq.~(\ref{eq:vBE})}
     \label{fig:kcplus2}
\end{figure}

As shown in Fig.~\ref{fig:kcminus}, the spin-band separation discussed above
for purely electric fields is a small effect even for very large fields . The
prospect of observing this effect is greatly improved for crossed electric and
magnetic fields. In this case, the term which mixes the $c+$ and $c-$ sectors
is given by Eq.~(\ref{eq:FSBE}) and the values of $h_3$ are significantly
larger than the corresponding terms $h_1$ and $h_2$ (see appendix B). As
mentioned previously, the case of crossed fields gives rise to an asymmetric
dispersion, and it leads to cross terms
$(\partial_x\theta_a)(\partial_x\phi_a)$ with coefficients proportional to the
difference $v_R-v_L$, which is quadratic in the external fields. The full
treatment of these and the cross term in Eq.~(\ref{eq:FSBE}) is beyond the
scope of the current paper. However, it is possible to get a sense of the
order of magnitudes involved by noting that in the present case the dominant
field effect is given by Eq.~(\ref{eq:FSBE}). We shall in the following study
the effect of this cross term by ignoring the asymmetry in the dispersion. The
relevant terms in the Hamiltonian are
\begin{eqnarray}
H&=&\frac{v_{c+}^0}{2}\int dx \Big[\frac{1}{K_{c+}^0}(\partial_x\theta_{c+})^2
+{K_{c+}^0}(\partial_x\phi_{c+})^2\Big]\nonumber\\
&&+\frac{v_{F}}{2}\int dx \Big[(\partial_x\theta_{c-})^2
+(\partial_x\phi_{c-})^2\Big]\nonumber\\
&&+g\int dx(\partial_x\theta_{c-})(\partial_x\phi_{c+}),
\end{eqnarray}
with $K_{c+}^0=v_F/v_{c+}^0$ defining the field-free values. In
diagonalizing the resulting Hamiltonian, care is required to ensure that
transformed fields respect commutation relations such as those of
Eq.~(\ref{eq:comm}). It is straightforward to show that the resulting plasmonic
modes retain linear dispersions having associated velocities and Luttinger
parameters given by
\begin{eqnarray}
v^2_{\pm} & = &\frac{v^2_F}{2} \left[ 1 + \left(\frac{ v_{c+}^0}{v_F}
\right)^2 \pm \sqrt{ \left( \frac{v_{c+}^0}{v_F}-1 \right)^2 + 4 g^2 \left(
\frac{v_{c+}^0}{v_F}
\right)^2} \right] \nonumber \\
K_{c\pm} & \equiv & v_F/v_{\pm}
\label{eq:vBE}
\end{eqnarray}
where $g$ is the coefficient of the $(\partial_x \theta_{c-})(\partial_x
\phi_{c+})$ term and is given by $g = 4e^2 h_3 U b / \pi \kappa$.  The
associated tunneling density of states into the bulk of the tube is given by
\begin{equation}
\alpha_{bulk} = \frac{1}{8} \left( \frac{1}{K_{c+}} + K_{c+} \right) + \frac{1}{16} (2 - K_{c+})\left(\frac{g}{v_F}\right)^2-\frac{1}{4}.
\end{equation}
This expression differs from the usual form (see the expressions below
Eq.~(\ref{eq:KcpE}) and below Eq.~(\ref{eq:KcpE}))
because of the unusual charge-current coupling term in the interactions.

Figure ~\ref{fig:kcminus2} shows the dependence of $K_{c-}$ on electric field
for a $(20,20)$ nanotube (with $K_{c+}^0 = 1/5$) in the presence of a 5 T magnetic field.
Interestingly, the fields render $K_{c-}$ larger than unity, reflecting
tendencies towards perfect conduction in the $c-$ sector. Furthermore, the
values attained by $K_{c-}$ show significant deviation from unity, making it
feasible to observe the proposed spin-charge-band separation for these field
configurations. Figure ~\ref{fig:kcplus2} shows a plot of $K_{c+}$ for the same situation.

\section{Luttinger liquid phases}
\label{sec:luttinger}
 The field-tuning of the Luttinger parameters discussed
in the previous section offers a viable way of tuning the groundstate of the
nanotube through different phases and different ordering tendencies. In the
absence of fields, Egger and Gogolin~\cite{egger} performed an involved
analysis using renormalization group arguments, refermionization and
considerations of various susceptbilities to predict the ordering tendencies
of the nanotube as a function of Luttinger parameters and temperature. In
particular, in the 'Luttinger liquid regime' which is the easiest to
experimentally access, wherein all four sectors $c\pm$ and $s\pm$, remain
ungapped, the prediction is that for the range of interaction values
$K_{c+}>1/5$, the system tends to show an inter-sublattice spin-density wave
(SDW$\pi$) ordering  while for $K_{c+}<1/5$ it tends to show inter-sublattice
charge-density wave (CDW$\pi$) ordering, where the corresponding operators are
defined, respectively, as
$\hat{O}_{CDW1}\sim\sum_{p\alpha\sigma}\psi^{\dagger}_{p\alpha\sigma}\psi_{-p\pm\alpha\sigma}$,
and
$\hat{O}_{SDW\pi}\sim\sum_{p\alpha\sigma}\sigma\psi^{\dagger}_{p\alpha\sigma}\psi_{-p\pm\alpha\sigma}$,
. These analyses involved considering operators associated with certain
orderings and determining the slowest decaying, equivalently, the most
relevant operator (i.e., of the smallest scaling dimension).

Here, we discuss the key changes that occur in the Luttinger regime in the
presence of fields. We focus on the Luttinger liquid regime and consider the
manner in which the field-tuned change in Luttinger parameters affect various
susceptibilities. We only consider the cases where either only an electric or
magnetic field is present; the cases when both fields are present are
extremely involved and beyond the scope of this paper. We do not take into
account the effect of non-quadratic bosonic terms generated by the fields;
even if relevant, we expect that the bare coupling associated with these terms
is so small that they only come into play at very low temperatures and not in
the Luttinger liquid regime.

{\bf Case of $\mathbf{B = 0}$, $\mathbf{E \neq 0}$ -} For the case of only an
electric field present, as discussed above, the effect of the field goes purely
into changing the value of $K_{c+}$. Given that experimentally the value of
$K_{c+}$ is around and oftentimes higher than $1/5$ and that the field tends to
reduce the value of $K_{c+}$, the electric field provides a unique means of
tuning from a tendency towards (SDW$\pi$) ordering to that of (CDW$\pi$)
ordering (see for example figure ~\ref{fig:kcplus}).

{\bf Case of $\mathbf{B \neq 0}$, $\mathbf{E = 0}$ -} The case of only a
magnetic field present, as discussed above, presents a slightly more complex
situation in which both $K_{c+}$ and $K_{c-}$ deviate from unity. As a result,
various susceptibilities acquire a $K_{c-}$ dependance in their scaling
behavior. For instance, operators associated with intra-sublattice ordering
such as
$\hat{O}_{CDW0}\sim\sum_{p\alpha\sigma}\psi^{\dagger}_{p\alpha\sigma}\psi_{p-\alpha\sigma}$
and
$\hat{O}_{SDW0}\sim\sum_{p\alpha\sigma}\sigma\psi^{\dagger}_{p\alpha\sigma}\psi_{p-\alpha\sigma}$,
which in the absence of fields are marginal, both acquire a scaling dimension
$(K_{c-}+K_{c-}^{-1}+2)/4$. Tendencies for superconducting order become
slightly stronger, though still irrelevant; the singlet pairing operator
$\hat{O}_{SC0}\sim\sum_{p\alpha\sigma}\sigma\psi_{p\alpha\sigma}\psi_{p-\alpha
-\sigma}$ acquires the scaling dimension $(K_{c-}+K_{c+}^{-1}+2)/4$.

To determine which ordering dominates, we consider the most relevant
candidates: the CDW$\pi$ and SDW$\pi$ operators, both of which have scaling
dimension $(K_{c-}+K_{c+}+2)/4$; parts of the second order CDW$\pi$ operator
denoted by $\hat{O}_{CDW1}^2$ that have scaling dimension $K_{c-}+K_{c+}$; and
a fourth order CDW$\pi$ operator denoted by $\hat{O}_{CDW\pi}^4$ which has
scaling dimensions $4K_{c+}$. Comparing these exponents shows that
$\hat{O}_{CDW\pi}$ and $\hat{O}_{SDW\pi}$ are more relevant than
$\hat{O}_{CDW\pi}^4$ for $15K_{c+}>2+K_{c-}$, a condition easier to satisfy in
the presence of fields than in the field-free case since $K_{c-}$ can then be
less than 1. Now $\hat{O}_{CDW\pi}^2$ is more relevant than
$\hat{O}_{CDW\pi}^4$ for $3K_{c+}>K_{c-}$. For $\hat{O}_{CDW\pi}^2$ to be more
relevant than $\hat{O}_{CDW\pi}$ and $\hat{O}_{SDW\pi}$  requires
$K_{c+}+K_{c-}<2/3$, a condition requiring inaccessibly strong interactions.
Finally, to determine whether $\hat{O}_{CDW\pi}$ or $\hat{O}_{SDW\pi}$
dominates, we appeal to the arguments of Ref.~\cite{egger}; at lower
temperatures where the physics is dominated by certain strong coupling fixed
points, pinning of the $\theta_{s+}$ mode suppresses $\hat{O}_{CDW\pi}$,
making its magnitude in the Luttinger phase smaller than that of
$\hat{O}_{SDW\pi}$. Though the methods of refermionization employed to reach
this conclusion are no longer valid for arbitrary values of $K_{c-}\neq 1$,
the strong coupling analysis still holds and we believe that a similar
conclusion can be reached for the finite magnetic field situation.

To summarize our results, while a detailed analysis, and considerations of
operators that are not taken into account in Ref. \cite{egger} are in order,
the most likely scenario is that the nanotube in the Luttinger regime for the
$B \neq 0$, $E = 0$ case is dominated by SDW$\pi$ ordering tendencies for
$15K_{c+}>2+K_{c-}$ (the more likely scenario gives that the deviation of
$K_{c-}$ from unity is not very large) and CDW$\pi$ ordering tendencies for
$15K_{c+}<2+K_{c-}$.

\section{Quantum dot physics}

For high resistance contacts or sufficiently low temperatures~\cite{giamarchi},
the nanotube shown in Fig.~\ref{fig:setup}  is only weakly coupled to the
leads, thus forming a quantum dot~\cite{QD}. The behavior of such dots and
related Coulomb blockade effects have been extensively studied by theory and
experiment~\cite{kbf,halperin}. Typical of quantum dot physics, Coulomb
blockade peaks have provided information on single-particle level spacings and
charging energies associated with the dot, as well as phase shifts due to
scattering at the edges of the nanotube dot. Moreover, under certain
conditions, the nanotube quantum dot has revealed a periodicity of four
associated with the degeneracy emerging from the band and spin degrees of
freedom. Recent work has also investigated the effects of a transverse magnetic
field on quantum dot behavior and the associated single-particle and charging
energies~\cite{bellucci2}. Here, we study the role of boundary conditions on
nanotube quantum dot physics, which requires subtle considerations in the
presence of fields. We then discuss the quantum dot behavior described by a
finite-sized version of the nanotube Luttinger liquid Hamiltonian which takes
into account relevant boundary conditions.

\subsection{Field dependent single-particle energy spectrum}

As is well known, finiteness of the tube length leads to a quantized single
particle spectrum which depends on the boundary conditions at the ends of the
tube. We assume that the wavefunctions at a given end are related by $\psi_{R
\alpha \sigma} = \sum_{\alpha' \sigma'} M_{\alpha \alpha' \sigma \sigma'}
\psi_{L \alpha' \sigma'}$ where $M$ is a matrix which depends on the the
microscopic details of the tube end but is assumed to be energy independent.
We specialize to the case that the boundary conditions do not affect spin;
that is we take $ M_{\alpha \alpha \sigma \sigma'} = S_{\alpha \alpha'}
\delta_{\sigma \sigma'}$, where $\delta_{\sigma\sigma'}$ is the Kronecker
delta function. We thus assume the absence of magnetic impurities and
local moments at the tube ends.

In order to obtain the appropriate boundary conditions for the case of an
asymmetric dispersion, we demand that the first quantized kinetic energy operator
$\hat{H}_0 = - i \hbar \sum_{r \alpha} r v_r \partial_x$ together with the
boundary conditions is self-adjoint~\cite{stone}. This treatment does not
account for the effect of interactions on the boundary conditions which would
be more naturally discussed in terms of the bosonic fields. Such an analysis
shows that there is an additional term in the current proportional to $g_2-g_4$
which vanishes for the density-density interaction considered here~\cite{fernandez}. By definition,
\begin{equation}
\langle \Psi, \hat{H}_0 \Psi \rangle = - i \hbar \sum_{r \alpha} \int dx \ r
v_{r} \psi^\dagger_{r \alpha} \partial_x \psi_{r \alpha}. \label{eq:1dKE}
\end{equation}
Since the boundary effects are assumed to be
independent of spin, we have dropped the spin index. For an arbitrary spinor
$\Psi$ with $\Psi = \left( \psi_{R} \ \psi_{L} \right)^T$ where $\psi_{R}$ and
$\psi_{L}$ are both two-component spinors in the Fermi point basis ($\psi_{r} =
\left( \psi_{r + } \  \psi_{r -} \right)^T$), self-adjointness gives
\begin{equation}
\langle \Psi, \hat{H}_0 \Psi \rangle = \langle \hat{H}_0 \Psi, \Psi \rangle.
\end{equation}
Integrating the left-hand side of this equation by parts gives
\begin{eqnarray*}
\langle \Psi, \hat{H}_0 \Psi \rangle &=& - i \sum_{\alpha = \pm} \int dx \left( v_R \psi^\ast_{R \alpha} \partial_x \psi_{R \alpha } - v_L \psi^\ast_{L \alpha } \partial_x \psi_{L \alpha } \right) \\
 &=& - i \sum_{\alpha = \pm} \bigg[ v_R \psi^\ast_{R \alpha} \psi_{R \alpha} - v_L \psi_{L \alpha}^\ast \psi_{L \alpha} \bigg]_{x = 0,L} \\ & &+  \langle \hat{H}_0 \Psi, \Psi \rangle
\end{eqnarray*}
Self-adjointness is satisfied as long as the boundary terms vanish, and this
leads to
\begin{equation}
\psi_{R \alpha} = S_{\alpha \alpha'} \psi_{L \alpha'}, \label{eq:bc}
\end{equation}
with $\sqrt{v_R/v_L}\, S$ unitary.

The details of the S-matrix can vary for each experimental set-up and depend
on physical attributes such as the substrate, the hardness of the confining
potential offered by the leads and the orientation of the tube's sublattices
with respect to the leads. These parameters can be incorporated as variables
in the boundary conditions which can then be utilized to obtain the
single-particle spectrum. The most general version of these boundary
conditions are outlined in Ref.~\cite{mccann} via the effective-mass model.

For a given S-matrix the spectrum of single particle states can be determined
by applying the condition of Eq.~(\ref{eq:bc}) at both ends and demanding that
both the left and right movers have the same energy. The two Fermi points give
rise to two sets of bands. The energy between two adjacent states in the
\emph{same} band is equal to $ \pi \hbar v_H / L$ where $v_H$ is the harmonic
mean,
\begin{equation}
v_H = \frac{2 v_R v_L}{v_R + v_L}.
\end{equation}
However, the energy offset of the bands from the Fermi energy depends on the
details of the S-matrix. In general, the two Fermi points will give rise to two
sets of energy states given by $ \pi n \hbar v_H / L + \Delta_1$ and $ \pi n
\hbar v_H /L + \Delta_2$ where $n \in \mathbb{Z}$.

Since we are ultimately interested in the spacing between Coulomb blockade
peaks, we focus on the energy difference between bands which we define as
$\Delta_{band} = \Delta_1 - \Delta_2$ (and for convenience we define
$\Delta_{band}$ such that $|\Delta_{band}| < \pi \hbar v_H / L$).
We examine two special cases for the S-matrix; deriving the $\Delta_{band}$
for the most general scattering matrix would be a straightforward extension.
First, consider the case in which the tube ends do not mix the Fermi points, though we allow the phase shift the electron suffers at the tube end $\delta_\pm (x)$ to be different for the two Fermi points ($\alpha = \pm$) and the two tube ends ($x = 0,L$). In this case we have $S_{++}(x) = \sqrt{v_L/v_R}
 e^{i \delta_+ (x)}$, $S_{--}(x) = \sqrt{v_L/v_R} e^{i \delta_-(x)}$ and
$S_{-+}(x) = S_{+-}(x) = 0$.
The energy offset between the bands takes the form
\begin{equation}
\Delta_{band} = \frac{ \pi \hbar v_H }{ L} {\cal F} \left[
\frac{\tilde{\delta}_1}{2 \pi}+\frac{2 t \Delta s}{\hbar \pi v_H / L}\right],
\label{eq:bandd}
\end{equation}
where $\tilde{\delta}_1 = \left( \delta_+(L) - \delta_+(0) \right) -
\left(\delta_-(L) - \delta_-(0) \right)$ and ${\cal F}(x) = x - \lfloor x
\rfloor$ and $\lfloor x \rfloor$ is the greatest integer less than or equal to
$x$, and the quantity $t \Delta s$ is the field induced offset between the two
Fermi points as defined in Eq.~(\ref{eq:mdiv}). (See Appendix~\ref{ap:3} for
derivation.)

Now, consider an electron that is completely scattered into the opposite Fermi
point at both boundaries. For simplicity we take
$ S_{\alpha \alpha'}(x) = \sqrt{v_L/v_R} e^{i \delta (x)} \delta_{\alpha,-\alpha'}$.
In this case, the splitting between bands takes the form
\begin{equation}
\Delta_{band} = \frac{ \pi \hbar v_H }{L} {\cal F} \left[
\frac{2}{\pi} k_F L + \frac{2 t \Delta s}{
\pi \hbar v_H / L} \left(\frac{v_R - v_L}{v_R + v_L} \right)\right].
\label{eq:bandod}
\end{equation}

For the limiting case of no fields (this also means that $v_H=v_R=v_L\equiv
v_F$), one expects the existence of sets of four single-particle states,
namely two degenerate sets of spin states and two sets of band states whose
energy splitting depends on the various phase shifts and the extent to which
modes at the two Fermi points mix. For no Fermi point mixing, the interband
splitting is $\hbar v_F \tilde{\delta}_1 / 2 L$ while for complete Fermi point
mixing, the splitting is $\frac{ \pi \hbar v_F }{L} {\cal F} [2 L k_F/\pi]$. Coulomb blockade experiments have
shown an interband band splitting of about 10\% \cite{bockrath} of the $\pi \hbar v_F/L$. Such a persistent approximate degeneracy in band energies for a range of tubes \cite{persistent} suggests that the magnitude of Fermi point mixing in these samples is minimal.

\subsection{Tunability of energy subband splitting}

As discussed above, the boundary conditions in a given experiment are not
directly observable since $\Delta_{band}$ depends on several parameters.
Fields provide a way of controlling $\Delta_{band}$ as well as studying its
physical origin in a particular sample. By scanning through various field
strengths, the variation of the band offset can reveal information about
the nature of boundary scattering. For example, the extent to which a given tube
interpolates between the two expressions given in Eqs.~(\ref{eq:bandd}) and
(\ref{eq:bandod}) can be used to determine the importance of (Fermi point) backscattering at the
tube ends. In the case of a \emph{natural} band degeneracy in a tube (that is,
no Fermi point mixing at the ends of the tube and $\tilde{\delta}_1 = 0$),
both electric and magnetic fields need to be applied to break the degeneracy;
the magnitude of the subband splitting as a function of fields can be
extracted from Eq. (\ref{eq:bandd}) by setting $\tilde{\delta}_1 = 0$. An
alternative approach for breaking the subband degeneracy was explored by
Ref.~\cite{halperin} in which a nonuniform external potential along the
tube was applied. However, this approach becomes infeasible for the case of a diagonal scattering matrix since it relies on band curvature away from half-filling.

Thus combining electric and magnetic fields can provide a means of breaking and
tuning the degeneracy of the quantum states of electrons inhabiting the
nanotube quantum dot. Of the four possible states discussed above, where the
direction of spin is defined with respect to the magnetic field, an extra
electron would occupy the ground state, which can be chosen to be any of the
four depending on the direction of the fields. The quantum state can be
characterized by a superspin inhabiting a $SU(2)\otimes SU(2)$ band and spin
space. The enhanced control of the spectrum of nanotubes that fields offer
would obviously have important implications for any potential quantum
information applications.

\subsection{Coulomb blockade physics}

We now consider quantum dot phenomena by taking into account interaction
effects in addition to the single-particle level spacing analysis of the
previous subsections. To present a coherent picture, we work within the context
of the Luttinger liquid description for field configurations that retain
gapless modes. This approach neglects the exchange energy within a dot, which
while shown to be present, is often much smaller in magnitude than the level
spacing and interaction energy \cite{bockrath}. While our
treatment captures salient features of quantum dot behavior, a full analysis of
the Luttinger liquid formulation for the most general boundary conditions is
yet to be performed.

Following the method of Ref.~\cite{kbf} (see also
Refs.~\cite{fabrizio,eggert}), for a finite sized version of
Eq.~(\ref{eq:H_tot}), we decompose the bosonic fields $\theta_{a}$ and
$\phi_{a}$ into sums of topological modes $\theta_{a}^0$, corresponding to a
net occupation number of the '$a$' sector $N_a= \frac{2}{\sqrt{\pi}} \int_0^L
\partial_x\theta_{a}^0 \ dx$ and harmonic modes corresponding to
plasmons. For simplicity, we consider the case of no Fermi point scattering so
that the Fermi point basis as defined by Eqs.~(\ref{eq:basisdef})
and~(\ref{eq:bosondensity}) and the band basis which diagonalizes the
boundary conditions in the previous subsection coincide (and therefore take $a
= c/s \pm$). Assuming the boundary conditions derived in the previous section,
we integrate out the $x$-dependence for the topological sector in the finite
sized version of Eq.~(\ref{eq:H_tot}). The resulting Hamiltonian associated
with 'charging energy' for each topological sector takes the form
\begin{eqnarray}
\label{eq:Ham_finite} {\cal H}_{a}
= \frac{1}{8} \epsilon_{a} N_{a}^2
\end{eqnarray}
where $\epsilon_{a} = \left( \frac{\hbar\pi v_H}{L} + 4 E_{a} \right)$ and
$E_{a}$ is equal to the interaction energy of a given mode.  The interaction
energy for the net charge sector comes from the forward scattering contribution
of Eq.~(\ref{eq:FS0}) to yield  $E_{c+} \approx \widetilde{V}(k) / L$. The
contribution due to the electric field given by Eq.~(\ref{eq:FSE}) is found to
be fourth order in fields and can thus be neglected. The magnetic field,
however, does contribute to the charging energy; from Eq.~(\ref{eq:FSB}), it
can be shown that $E_{c-} = 2 e^2 h_2 b^2 / \kappa L$. This expression
represents an upper bound that assumes the limit of no Fermi-point mixing.

The topological modes correspond to the addition of charges onto the dot; in
addition, plasmon modes that correspond to harmonic vibrations of the densities
in the various sectors are present. In principle, the procedure employed in
Ref.~\cite{kbf,fabrizio} to derive the structure of these plasmonic excitations
can be generalized to the case of the asymmetric dispersion by incorporating
the asymmetric description in Ref.~\cite{trushin}. Here we forego such a
derivation; most quantum dot experiments involving adiabatic tuning of
parameters such as gate voltage and thus probe purely ground state properties
determined by the topological sectors.

In the presence of a gate voltage $V_G$, the Hamiltonian associated with the
topological modes of the nanotube quantum dot is given by
\begin{equation}
H_L=\sum_{a=\pm c/s}{\cal H}_{a} - \mu N_{c +}+\frac{1}{2}\Delta_{band}
N_{c-}-\Delta_Z N_{s+}, \label{eq:Ham_qtmdot}
\end{equation}
where $\mu$ is essentially $e V_G$ and the term $\Delta_Z = \mu_B B$ accounts
for the Zeeman splitting. The hierarchy in energy scales can be summarized for
a typical tube length of $L = 500$ nm. The intrasubband splitting is $3.3$
meV. Therefore we have that $\epsilon_{c-} \approx \epsilon_{s+/-} = 3.3$ meV.
The charging energy is then $\epsilon_{c+} = 47.7$ meV. Consider a tube for
which the ends of the tube do not appreciably change the Fermi point. In that
case the sign of $\Delta_{band}$ can be changed by reasonable electric and
magnetic field values (for example, this is true for an $n = 15$ tube, a 6 T
magnetic field and an electric field of order 1 V/nm). Finally, $\mu_B \approx
0.058$ meV/T and therefore for most situations the Zeeman splitting will be at
least an order of magnitude smaller than the other effects considered here.

The Hamiltonian in Eq.~(\ref{eq:Ham_qtmdot}) forms a starting point for
analyzing quantum dot and Coulomb blockade behavior in short nanotubes.
Typically, conductance is measured across the tube as a function of the
applied gate voltage $V_G$ and a bias voltage $V_B$; while for the most part,
energetic costs impede the flow of electrons, at special degeneracy points
that equally favor an occupation of $N$ and $N+1$ electrons, zero bias Coulomb
blockade conductance peaks can be observed. Given that the occupation numbers
$N_{a}$ with $a = \{c/s,\pm\}$ are good quantum numbers, the net energy of the
system $E_L=\langle H_L\rangle$ for a given configuration of electrons
corresponds to a given combination of eigenvalues of $N_{a}$.  The equilibrium
configuration of electrons on the quantum dot can thus be derived by
minimization, i.e., by requiring $\frac{\partial E_L}{\partial N_{a}} = 0$ for
all $N_{a}$ sectors, subject to the physical constraint that that electron
occupation numbers $N_{\alpha=\pm,\sigma=\uparrow/\downarrow}$ take on integer
values. The relationship between these two bases is given by $N_{c \pm} =
(N_{+\uparrow} + N_{+\downarrow})\pm (N_{-\uparrow} + N_{-\downarrow}) $ and
$N_{s \pm} = (N_{+\uparrow} - N_{+\downarrow})\pm (N_{-\uparrow} -
N_{-\downarrow})$.

As an illustrative example, consider the Coulomb blockade situation for the
simple case of no fields, no band or Zeeman splitting
($\Delta_{band}=\Delta_z=0$ and only the charging energy in the $c+$ sector is
non-zero). Applying the condition that $\frac{\partial E_L}{\partial N_{a}} =
0$ gives $\epsilon_{c+}N_{c+} - 4 \mu = 0$ and $N_{c-} = N_{s+} = N_{s-} = 0$
(where $\mu$ is varied by $V_G$). These conditions suggest that all the states
at a given energy level will fill before filling the subsequent energy level.
Now, as the chemical potential is increased, the first extra electron added
onto the dot can occupy any of the degenerate states of the
$|\alpha=\pm,\sigma=\uparrow/\downarrow\rangle$ space. Suppose that this
electron goes in the $\alpha=+$ band with its spin up. The state of the tube
is then characterized by the quantum numbers $(N_{c+},N_{c-},N_{s+},N_{s-}) =
(1,1,1,1)$. Further increasing the chemical potential by an amount $E_{c+}$
will add the next electron to any of the remaining three states. For example
if the filling obeys Hund's rule (i.e., assuming the exchange interaction
which we have thus far neglected) then band $\alpha=-$ would be filled with
spin up electron thus giving the state $(2,0,2,0)$. An additional third
electron can occupy $+\downarrow$ or $-\downarrow$ state. The energy cost for
adding each of these extra electrons is $E_{c+}$, reflecting the Coulomb
charging energy. However, the fifth extra electron requires a chemical
potential increase of $\frac{\pi \hbar v_H}{L} + E_{c+}$, reflecting the
Coulomb energy as well as the excitation energy required to occupy the next
energy level. This analysis, executed within the Luttinger liquid description,
reproduces the periodicity of four observed in experiment \cite{bockrath}.

In the presence of fields and intrinsic subband splitting, an analysis similar
to the one above can be performed for altered values of $\epsilon$ and $K_a$'s
and the orders of magnitude discussed after Eq.~(\ref{eq:Ham_qtmdot}). We take
$\Delta_Z \approx 0$ and $\Delta_{band} > 0$. Furthermore, suppose that for a
given tube that a nonzero magnetic field gave $E_{c-} < \Delta_{band}/2$. In
that case, the tendency to minimize $N_{c-}$ would give rise to a shell-filling
opposite to the usual Hund-like filling. For example, the order in which states
are filled could take te forn $-\uparrow, -\downarrow, +\uparrow, +\downarrow$
or equivalently $(1,-1,1,-1) \rightarrow (2,-2,0,0) \rightarrow (3,-1,1,1)
\rightarrow (4,0,0,0)$. Instead of the Coulomb blockade peaks being of
periodicity 4 described above (spaced apart by $E_{c+}, E_{c+}, E_{c+},
E_{c+}+{\pi \hbar v_H}/{L}$, spacings become $E_{c+}+E_{c-},
E_{c+}-3E_{c-}+\Delta_{band}, E_{c+}+E_{c-}, E_{c+}+E_{c-}+{\pi
\hbar v_H}/{L}-\Delta_{band}$.

The effect of shell-filling in nanotube quantum dots has been investigated
experimentally by Liang et al.~\cite{bockrath}. The four-electron periodicity
was clearly observed via transport measurement. Parameters of charging energy
and exchange energy were determined. The above results, including
field-dependent Luttinger parameters, the effect of boundary conditions, the
band splitting, asymmetric dispersions and additional two-electron
periodicity can thus be studied in a similar setup with transverse fields.
Additionally, the Fabry-Perot transmission resonances in the
presence of a transverse magnetic field predicted by Bellucci and
Onorato~\cite{bellucci2}, as well as possible resonances by both transverse
electric and magnetic fields (and their relative angles), can be investigated
in the same experimental setup. While our arguments here have been confined to
adiabatic tuning and zero-bias conductance, our approach can be used to
investigate non-equilibrium phenomena, temperature dependences, higher order
tunneling events such as cotunneling and non-adiabatic tuning. Each of these
considerations, which is beyond the scope of this paper, would involve excitations of the plasmonic modes.

\section{Discussion; Relevance to Experiment}
 We have investigated the effects of transverse electric and magnetic fields
 on armchair carbon nanotubes. We found that fields can break
 several symmetries inherent to the carbon
nanotubes---the valley degeneracy, the left-right-mover degeneracy, and the
particle-hole symmetry.  The magnitude of a gap in the nanotube spectrum can be
continuously tuned by varying the strength and the relative angle of the two
fields. We also found that the electron-electron interaction is modified by
both fields and thus Luttinger-liquid parameters can be tuned by fields. In
particular, an interesting consequence is the possibility of spin-charge-band
separation. We also discussed how the fields can be used to study boundary
effects in finite sized tubes and to describe the Coulomb-blockade physics in
presence of fields.

 Each of these salient features can become manifest in experiment, some in dramatic
 ways. While we summarize these experimental signatures here, details of the physics
 and orders of magnitude can be found in the relevant section. At the band structure level,
 the reduction of the Fermi velocity can be observed by measuring the particle level
 spacing in a finite size tube. The shift in the Fermi momentum induced
 by either field may be detected by virtue of the associated Friedel oscillations
 around a dopant or impurity using a STM (see for example \cite{sprunger})
   For both fields present at
  an arbitrary angle to one another, a continuous conduction gap occurs at the Fermi
  energy, discernible via direct conductance measurements, shifts in Coulomb
  blockade peaks and STM measurements~\cite{saito,coskun,dekker}.  In the transport measurement, the conductance peak
  can vary from $0$ to $2e^2/h$, and to $4 e^2/h$, depending on the fields and the chemical
potential. Perhaps the most dramatic prediction of
  band structure analysis is that an electric and a magnetic field
  perpendicular to one another and the tube axis would give rise to a current
  carrying ground state. Thus, a tube subject to this field configuration placed
  across two leads should induce a measurable current even in the absence of an applied
  voltage drop across the leads.

  In the regime in which SWNTs exhibit Luttinger liquid behavior, strong
  enough fields can give rise to significant changes. The value of the Luttinger
  parameter associated with the net charge sector can be tuned via either an
  electric field or a magnetic field or both. Furthermore, the presence of an electric
  field gives rise to density-density interactions associated with the
  difference in densities in the two bands and results in the deviation of the
  associated Luttinger parameter $K_{c-}$ from its non-interacting value of
  unity. The tunneling density-of-states, a quantity ubiquitous to a range of
experiments including scanning tunneling microscope studies and conductance
measurements \cite{giamarchi,saito,mceuen2,yao}, would reflect these changes in its power-law
behavior.

 As discussed in Sec.\ref{sec:luttinger}, changing the Luttinger parameter via
fields can result in tuning through phases having
charge-density-wave order or spin-density-wave order; such phases are potentially measurable
in STM and neutron scattering experiments. Another exciting prospect comes
about in the tuning of $K_{c-}$, namely, that of spin-band-charge separation.
In the past, momentum resolved tunneling experiments have resolved charge and
spin modes moving at different velocities in quantum wires~\cite{Yacoby05}; in
principle, a more elaborate version of such an experiment could detect charge,
spin and band modes moving at three different velocities in nanotubes.

 In the quantum dot regime, the application of fields acts as a controlled
 means of changing the Coulomb blockade structure of the dot, and could potentially
 have a plethora of applications. For tubes that
 preserve the four-fold degeneracy emerging from spin and band degrees of
 freedom in the absence of fields, the presence of fields can serve to
 break this degeneracy. For tubes that show a lack of degeneracy, fields
 provide a way of determining the origin of degeneracy breaking. This effect has
 potential applications to quantum information. Through achieving a desired amount of
 degeneracy breaking for the four states, fields can be an effective means of
 initializing the quantum state of an extra electron added on to the dot.
 Having initialized a quantum state, transitions can be induced to other states. Additionally, as has been demonstrated
for semi-conducting quantum dot spin states~\cite{kouwenhoven}, superpositions can be created by applying oscillating fields. For the energies quoted above, oscillation frequencies would be on the order of $10^{11}$ Hz.

To conclude, transverse fields induce a rich range of physical effects in the
electronic properties of SWNTs from band structure effects to one-dimensional
behavior to quantum dot physics. Several of these features are very much within
the reach of current experimental capabilities and are of both fundamental and
applied value.

We would like to acknowledge M. Stone for illuminating discussions. One of us
(WD) would like to thank Sarang Gopalakrishnan for discussions on the various
symmetries of the band structure and Jeremy McMinis for his
assistance in preparing the nanotube figure (Fig. 1). This work was
supported by the NSF under the grant DMR-0605813 (SV,WD). TCW was supported by
IQC, NSERC and ORF.

\appendix

\section{Band structure calculation}
\label{app:band}
 In the absence of any fields, the eigenstates of an
infinitely long armchair nanotube are superpositions of the states $|
\Phi_{A/B} \ell \rangle$ defined by Eq. (1). For the particular case of an
armchair nanotube Eq. (4) gives
\begin{widetext}
\begin{eqnarray*}
\langle \Phi^{\ell'}_A \left| H_B \right| \Phi^{\ell}_B \rangle = -\frac{t}{N} \sum_s e^{i 2 \pi \frac{(\ell - \ell')s}{L}} \bigg\{ e^{ \frac{i 2 \pi \ell a}{\sqrt{3}L}} + 2 e^{-\frac{i 2 \pi \ell a}{2 \sqrt{3} L}}  \cos \left( \frac{k_y a}{2} + \sqrt{3} B \frac{|e|}{\hbar} \left( \frac{L}{2 \pi} \right)^2 \left[\cos{\frac{2 \pi}{L} s} - \cos \frac{2 \pi}{L} \left(s-\frac{a}{2 \sqrt{3}} \right) \right] \right) \bigg\}.
\end{eqnarray*}
\end{widetext}
where $b = \frac{\sqrt{3} B |e| L^2}{4 \pi^2 \hbar}$. For small magnetic fields ($b \ll 1$) we have
\begin{equation}
\langle \Phi^{\ell'}_A \left| H_B \right| \Phi^{\ell}_B \rangle = t b \sin{\frac{k a}{2}} e^{\frac{i \pi \ell}{3n}}\left(1-e^{\pm \frac{i \pi}{3n}}\right)
\end{equation}
for $\ell - \ell' = \pm 1 \mbox{ mod }N$ and
\begin{equation}
\langle \Phi^{0}_A \left| H_B \right| \Phi^{0}_B \rangle = t b^2 \cos{\frac{k a}{2}} \left(1-\cos{\frac{\pi}{3n}}\right).
\end{equation}

Applying perturbation theory near the Fermi points requires care because of the near degeneracy of the left and right movers. In this case, the usual procedure of first diagonalizing the nearly degenerate subspace fails because the matrix elements within this subspace vanish to the order we are working. However, there are matrix elements for transitions to other energy levels ($\ell \neq 0$) and these matrix elements give rise to an effective interaction between states in the nearly degenerate subspace. The subspace can be diagonalized once these additional interactions are taken into account \cite{landau}.

For the gapless case, the low energy spectrum near half-filling is
\begin{equation}
\label{eq:mdiv2} \epsilon_{r \alpha} (k) =  \hbar r \tilde{v}_{F,r} \left( k -
\alpha \tilde{k}_F \right) + \alpha t \Delta s + \mathcal{O}(k^2),
\end{equation}The renormalized Fermi velocity is given by
\begin{subequations}
\begin{equation}
v_{r} = v_F \left(1 - \Delta v_1 b^2 - \Delta v_2 U_y^2 \pm \Delta v_3 b U_y \right),
\end{equation}
where
\begin{equation}
\Delta v_1 = \frac{5 + 4 \cos{\frac{\pi}{n}}}{3 \left(1 + 2 \cos{\frac{\pi}{3n}} \right)^2}, \label{eq:v1}
\end{equation}
\begin{equation}
\Delta v_2 = \frac{3 + \cos{\frac{\pi}{3n}} + 2 \cos{\frac{2 \pi}{3n}}}{12 \left(1 - \cos{\frac{\pi}{n}} \right)} \label{eq:v2}
\end{equation}
\begin{equation}
\Delta v_3 = \frac{\left(\cos{\frac{\pi}{6n}} + \cos{\frac{5 \pi}{6n}} \right) \csc{\frac{\pi}{6n}}}{\sqrt{3} \left(1 + 2 \cos{\frac{\pi}{3n}}\right)^2}. \label{eq:v3}
\end{equation}
\end{subequations}
and
\begin{equation}
\tilde{k}_F = \left[ \frac{4 \pi}{3 a} + \frac{U_y^2}{2 \sqrt{3} \left( 1 + 2 \cos \frac{\pi}{3n} \right)}+\frac{8}{\sqrt{3}} \sin^2{\left(\frac{\pi}{6n}\right)} b^2 \right] \label{eq:kf}
\end{equation}
The shift between the two Fermi points (see Eq. (10)) is given by
\begin{equation}
\Delta s = \frac{\sqrt{3} \sin \frac{\pi}{3n}}{1 + 2 \cos{\frac{\pi}{3n}}} b U_y. \label{eq:ds}
\end{equation}
For the case of mutually perpendicular fields discussed in section II, the electronic densities are given by the vector
\begin{equation}
\label{eq:rhora} \rho_{r \alpha}(s) = \frac{1}{2} \left( \begin{array}{ccc}
1 + g_1 r u \cos\left(\frac{s}{R} + \frac{\pi}{6n}\right) - g_2 b \alpha \cos\left(\frac{s}{R} + \frac{\pi}{3n} \right)  \\
1 - g_1 r u \cos\left(\frac{s}{R} - \frac{\pi}{6n}\right) + g_2 b \alpha
\cos\left(\frac{s}{R} - \frac{\pi}{3n} \right)  \end{array} \right),
\end{equation}
where the upper and lower components are the electronic densities over the $A$ and $B$ sublattices respectively. The constants $g_1$ and $g_2$ are given by
\begin{eqnarray}
g_1 &=& \frac{1}{2} \csc \frac{\pi}{2n} \\
g_2 &=& \sqrt{3} \csc \frac{\pi}{2n}.
\end{eqnarray}

\section{Interaction terms}
\label{ap:2}
In order to find the form of the effective interaction $V^{r r'}_{\alpha \alpha'}$ we need account for not only the radial dependence of the wave functions but also for the physical separation between the sublattices. Although the factorization performed in Eq.~\ref{eq:wvfn} is an approximation, we may still account for the physical separation of the sublattices. We follow the approach introduced in \cite{egger}.
\begin{eqnarray}
V^{rr'}_{\alpha \alpha'}(x-x') &=& \int_0^{2\pi R} \int_0^{2\pi R} \frac{ds \ ds'}{(2 \pi R)^2} \times \nonumber  \\
& & \rho_{r \alpha}^T(s) \left( \begin{array}{cc} U(0) & U(a_{c})\nonumber \\ U(-a_{c}) & U(0)  \label{eq:Vrr} \\ \end{array} \right)  \rho_{r' \alpha'}(s')
\end{eqnarray}
where $U(d)$ is a shorthand for the Coulomb interaction with an offset $d$, that is
\begin{equation}
U(d) = U(x-x',s-s'+d)
\end{equation}
where the right hand side of this equation is given by Eq ~(\ref{eq:coulomb}).

The constant $c_0$ which appears in Eq. ~(\ref{eq:one}) is given by
\begin{eqnarray*}
\label{eq:c} c_0(n) = -\gamma &-& \frac{1}{4 \pi} \int_0^{2 \pi} d \varphi \,
\ln\big[\cos^2\frac{\phi}{2}+\frac{\pi^2}{3n^2}\big].
\end{eqnarray*}

Similarly we find that the values of $h_1$ and $h_2$ defined in section 3 are given by
\begin{eqnarray*}
h_1(n) = \left(c_2^2 - c_1^2 \right) f_1(n) + 2 c_1 c_2 f_2(n) + 2\left(c_1^2 + c_2^2\right) f_3(n),
\end{eqnarray*}
\begin{eqnarray*}
h_2(n) =  \left(c_4^2 - c_3^2 \right) f_1(n) + 2 c_3 c_4 f_2(n) + 2\left(c_3^2 + c_4^2\right) f_3(n),
\end{eqnarray*}
and
\begin{eqnarray*}
h_3(n) = 2 [ \left( c_1 c_3 + c_2 c_4 \right) f_1(n) + \left(c_1 c_4 + c_2 c_3 \right) f_2(n) \\
+ 2 (c_1 c_3 + c_2 c_4) f_3(n)  ],
\end{eqnarray*}
where
\begin{eqnarray*}
f_1(n) = & &\int_{-\pi R}^{\pi R} \frac{dz}{2 \pi R} \left\{ \ln
\left[ \cos^2 \frac{z-a_{c}}{2R} + \left(\frac{a_z}{2R} \right)^2 \right] \right. \\
 &+&  \left. \ln \left[ \cos^2 \frac{z+a_{c}}{2R} + \left( \frac{a_z}{2R} \right)^2 \right] \right\}  \cos \frac{z}{R},
\end{eqnarray*}
\begin{eqnarray*}
f_2(n) = & &\int_{-\pi R}^{\pi R} \frac{dz}{2 \pi R} \left\{ \ln
\left[ \cos^2 \frac{z-a_{c}}{2R} + \left(\frac{a_z}{2R} \right)^2 \right] \right.\\
 &-&  \left. \ln \left[ \cos^2 \frac{z+a_{c}}{2R} + \left( \frac{a_z}{2R} \right)^2 \right] \right\}  \sin \frac{z}{R},
\end{eqnarray*}
\begin{eqnarray*}
f_3(n) = \int_{-\pi R}^{\pi R} \frac{dz}{2 \pi R} \ln \left[ \cos^2 \frac{z}{2R} + \left( \frac{a_z}{2R} \right)^2 \right]
\end{eqnarray*}
and
\begin{eqnarray*}
c_1 &=& \frac{1}{2} \csc \frac{\pi}{2n} \cos \frac{\pi}{6n} \\
c_2 &=& \frac{1}{2} \csc \frac{\pi}{2n} \sin \frac{\pi}{6n} \\
c_3 &=& \sqrt{3} \csc \frac{\pi}{2n} \sin \frac{\pi}{6n} \cos \frac{\pi}{3n} \\
c_4 &=& \sqrt{3} \csc \frac{\pi}{2n} \sin \frac{\pi}{6n} \sin \frac{\pi}{3n}. \\
\end{eqnarray*}
The values of $c_0$, $h_1$, $h_2$ and $h_3$ have been tabulated for various tube sizes and are displayed below.
\begin{center}
\begin{tabular}{|r|r|r|r|r|}
\hline
$n$  & $c_0$ & $h_1$ & $h_2$ & $h_3$  \\
\hline
5  & -0.239   &  0.491    & 0.112  & 3.780    \\
10 & -0.064   &  0.697    & 0.040  & 10.370   \\
15 &  0.005   &  0.786    & 0.020  & 17.417   \\
20 &  0.025   &  0.834    & 0.122  & 24.605   \\
25 &  0.434   &  0.865    & 0.008  & 31.855   \\
\hline
\end{tabular}
\end{center}
\section{Single-particle spectrum with boundary scattering}
\label{ap:3} Here we illustrate the case in which the tube ends do not mix
Fermi points; that is $S_{++}(x) = \sqrt{v_L/v_R}e^{i \delta_+ (x)}$,
$S_{--}(x) = \sqrt{v_L/v_R} e^{i \delta_-(x)}$ and $S_{-+}(x) = S_{+-}(x) = 0$. Assume the
wavefunctions are
\begin{eqnarray}
\psi_R=\begin{pmatrix} A e^{i k_{R+} x} \cr B e^{i k_{R-} x}
\end{pmatrix},
\\
\psi_L=\begin{pmatrix} C e^{i k_{L+} x} \cr D e^{i k_{L-} x}
\end{pmatrix}.
\end{eqnarray}
The energy spectrum for each branch is described by Eq.~(\ref{eq:mdiv}), e.g.,
$\epsilon_{R,\pm}=\hbar v_R(k_{R\pm}\mp k_F)\pm t \Delta s$. The
self-adjointness condition~(\ref{eq:bc}) at $x=0, L$ gives
\begin{eqnarray*}
\!\!\!\!\!\!\!\!\!\!\!\!\!\!\!\!\!\!\!\!&&\begin{pmatrix} A  \cr B
\end{pmatrix}=\sqrt{\frac{v_L}{v_R}}
\begin{pmatrix} e^{i\delta_+(0)}& 0\cr
0& e^{i\delta_-(0)}
\end{pmatrix}\begin{pmatrix} C  \cr D
\end{pmatrix}\\
\!\!\!\!\!\!\!\!\!\!\!\!\!\!\!\!\!\!\!\!&&\begin{pmatrix} A e^{ik_{R+}L} \cr B
e^{ik_{R-}L}
\end{pmatrix}=\sqrt{\frac{v_L}{v_R}}
\begin{pmatrix} e^{i\delta_+(L)}& 0\cr
0& e^{i\delta_-(L)}
\end{pmatrix}\begin{pmatrix} C e^{ik_{L+}L}  \cr D e^{ik_{L-}L}
\end{pmatrix} \\
\end{eqnarray*}
This gives constraints on the four momenta
\begin{eqnarray}
(k_{R\pm}-k_{L\pm})L=2\pi n_\pm+\delta_\pm(L)-\delta_\pm(0),
\end{eqnarray}
where $n_\pm$ are arbitrary integers. For the $\alpha=+$ branch, the energy
levels corresponding to $k_{R+}$ and $k_{L+}$ should be equal (in order for the
wavefunction to represent an eigenstate). Hence
\begin{eqnarray}
\hbar v_R(k_{R+}-k_F)+t\Delta s=-\hbar v_L(k_{L+}-k_F)+t\Delta s,
\end{eqnarray}
which gives
\begin{equation}
k_{L+}= -k_F-v_R\frac{2\pi n_++\delta_+(L)-\delta_+(0)}{(v_R+v_L)L},
\end{equation}
and thus
\begin{equation}
\epsilon_+(n_+)=\hbar v_H\frac{2\pi n_++\delta_+(L)-\delta_+(0)}{2L}+\Delta s,
\end{equation}
where $v_H\equiv 2v_Rv_L/(v_R+v_L)$. Similarly, for $\alpha=-$, we have
\begin{equation}
\epsilon_-(n_-)=\hbar v_H\frac{2\pi n_-+\delta_-(L)-\delta_-(0)}{2L}-\Delta s.
\end{equation}
Hence, we arrive at the interband energy difference
\begin{equation*}
\Delta_{band}=\epsilon_+(n_+) - \epsilon_-(n-) = \frac{\pi\hbar
v_H}{L}\Big(\tilde{n}+\frac{\tilde\delta_1}{2\pi}+\frac{2t\Delta s}{\hbar\pi
v_H/L}\Big),
\end{equation*}
where $\tilde{\delta}_1 = \Big( \delta_+(L) - \delta_+(0) \Big) -
\Big(\delta_-(L) - \delta_-(0) \Big)$ and the $\tilde{n}_\pm$ are selected so that $|\Delta_{band}|$ is less than the intraband spacing. Hence we obtain
Eq.~(\ref{eq:bandd}). A similar consideration leads to Eq.~(\ref{eq:bandod}).

\end{document}